\documentclass[preprint]{aastex63}

\usepackage[utf8]{inputenc} 
\usepackage{hyperref}       
\usepackage{url}            
\usepackage{booktabs}       
\usepackage{amsfonts}       
\usepackage{nicefrac}       
\usepackage{microtype}      
\usepackage{xcolor}         

\usepackage{capt-of}
\usepackage{sidecap}
\usepackage{amsmath}
\usepackage{amssymb}
\usepackage{tikz}
\usetikzlibrary{shapes}
\usetikzlibrary{calc}
\usetikzlibrary{positioning}
\usepackage{pgf}
\usepackage{pgfplots}
\usepackage{pgfplotstable}

\newcommand{\module}[1]{\textit{#1}}

\begin{document}

\title{Predicting the Thermal Sunyaev-Zel'dovich Field using Modular and Equivariant Set-Based Neural Networks}

\author{Leander Thiele}
\affiliation{Department of Physics, Princeton University, Jadwin Hall,
Princeton NJ 08544, USA}

\author{Miles Cranmer}
\affiliation{Department of Astrophysical Sciences, Princeton University, Peyton Hall,
Princeton NJ 08544, USA}

\author{William Coulton}
\affiliation{Center for Computational Astrophysics, Flatiron Institute,
162 5th Ave, New York NY 10010, USA}

\author{Shirley Ho}
\affiliation{Center for Computational Astrophysics, Flatiron Institute,
162 5th Ave, New York NY 10010, USA}
\affiliation{Department of Astrophysical Sciences, Princeton University, Peyton Hall,
Princeton NJ 08544, USA}
\affiliation{Department of Physics, Carnegie Mellon University,
Pittsburgh PA 15213, USA}

\author{David N. Spergel}
\affiliation{Center for Computational Astrophysics, Flatiron Institute,
162 5th Ave, New York NY 10010, USA}
\affiliation{Department of Astrophysical Sciences, Princeton University, Peyton Hall,
Princeton NJ 08544, USA}


\begin{abstract}
Theoretical uncertainty limits our ability to extract cosmological information
from baryonic fields such as the thermal Sunyaev-Zel'dovich (tSZ) effect.
Being sourced by the electron pressure field, the tSZ effect depends on baryonic physics
that is usually modeled by expensive hydrodynamic simulations.
We train neural networks on the IllustrisTNG-300 cosmological simulation
to predict the continuous electron pressure field in galaxy clusters from gravity-only simulations.
Modeling clusters is challenging for neural networks as most of the gas pressure
is concentrated in a handful of voxels and even the largest hydrodynamical simulations
contain only a few hundred clusters that can be used for training.
Instead of conventional convolutional neural net (CNN) architectures,
we choose to employ a rotationally equivariant DeepSets architecture
to operate directly on the set of dark matter particles.
We argue that set-based architectures provide distinct advantages over CNNs.
For example, we can enforce exact rotational and permutation equivariance,
incorporate existing knowledge on the tSZ field,
and work with sparse fields as are standard in cosmology.
We compose our architecture with separate, physically meaningful modules,
making it amenable to interpretation.
For example, we can separately study the influence of local and cluster-scale environment,
determine that cluster triaxiality has negligible impact,
and train a module that corrects for mis-centering.
Our model improves by $70\,\%$ on analytic profiles
fit to the same simulation data.
We argue that the electron pressure field, viewed as a function of a gravity-only simulation,
has inherent stochasticity,
and model this property through a conditional-VAE extension to the network.
This modification yields further improvement by $7\,\%$,
it is limited by our small training set however.
We envision that our method 
will prove useful in problems beyond the specific one considered here.\footnote{
We make our code publicly available at
\href{https://github.com/leanderthiele/tSZ_DeepSet}{this URL}.
Data products, trained models, and hyperparameter SQL databases will be shared upon reasonable request.}
\end{abstract}

\section{Introduction}
\label{sec:intro}

A pressing problem in cosmology is the accurate modeling of observables sourced
or influenced by physics beyond gravity, in short, baryonic effects.
Hydrodynamic simulations are the canonical forward model for such fields;
however, their computational cost is too high for them to be a viable contender in generating
the vast number of realizations necessary to sample distributions.
Thus, an approach that has recently emerged is the use of neural networks to map cheaper gravity-only
simulations to their full-physics counterparts.
Not only does this idea enable a substantial speed-up in generating realizations,
but it could also improve our physical understanding; to this aim interpretable models are required.

One example of a baryonic observable that we aim to model is the thermal
Sunyaev-Zel'dovich (tSZ) effect \citep{Zeldovich1969,Sunyaev1970,Sunyaev1980,Carlstrom2002,Gatti2021}.
The tSZ field is a secondary cosmic microwave background (CMB) anisotropy generated by inverse Compton
scattering between electrons in the Universe's large scale structure and CMB photons.
Its intensity at sky position $\hat n$, to leading order determined by the Compton-$y$ parameter,
is given by a line-of-sight integral over the electron pressure, $P_e(\vec x)$:
\begin{equation}
y(\hat n) = \frac{\sigma_T}{m_e c^2} \int dl\,P_e(\hat n, l)\,.
\end{equation}

The problem that we tackle in this work is the prediction of the electron pressure $P_e(\vec x)$
given a gravity-only simulation.
Since this is a translationally equivariant spatial problem, the seemingly natural approach
chosen for similar problems,
e.g., by \cite{Troester2019,Yip2019,Zhang2019,Kasmanoff2020,Thiele2020,Wadekar2021,Rothschild2021},
is a convolutional neural net (CNN), taking as input the density field of a gravity-only simulation.
However, in this work we argue that existing domain knowledge on $P_e(\vec x)$ and similar fields
renders the CNN approach inferior to a set-based architecture.
In fact, electron pressure values high enough to affect observables are predominantly found in
massive gravitationally collapsed structures, called clusters.
A neural network should naturally take this property into account.
From this point of view, translational invariance is in fact broken, negating the main advantage of CNNs.

Besides the heavy bias towards clusters, the electron pressure field exhibits a related feature that
further inhibits the performance of CNNs.
Even within a cluster the mass and electron pressure is highly concentrated towards the center.
Thus, CNNs operating on the density field are highly inefficient in that they need to maintain a uniformly
high spatial resolution in order to resolve the important small-scale details near the cluster center,
while conversely this high resolution leads to a waste of resources in the low-density outer regions.

To a first approximation, clusters are described by their mass, $M_{200}$, and radius, $R_{200}$.
Here, we employ the standard convention of mass within a spherical volume that contains an average
density 200 times larger than the mean density of the Universe.
These quantities determine a characteristic pressure scale: $P_{200} \propto M_{200}/R_{200}$.
The electron pressure\footnote{We use $\vec r$ for coordinates relative to a cluster's position.} $P_e(\vec r)$
is to leading order a spherically symmetric function,
commonly approximated as a generalized~\mbox{Navarro-Frenk-White}~(GNFW)
profile~\citep{NFW1997,Nagai2007,Battaglia2012},
\begin{equation}
	P_e(\vec r) \approx \text{GNFW}(|\vec r|; M_{200}, R_{200})\,,
\label{eq:GNFW}
\end{equation}
which we will use as the benchmark
(with the parameterization as chosen in \cite{Battaglia2012}, fitted to our data).
There is an inherent random element in the electron pressure field if viewed
as a function of a gravity-only simulation's snapshot at a given time.
The reason is that chaos washes out some of the history;
in particular the time-integrated activity of the active galactic nuclei (AGNs)
is difficult to infer.

We propose to learn a probabilistic mapping directly from the simulation representation,
i.e., from a set of dark matter particles with associated positions $\vec q_i$
and velocities $\vec v_i$.
For a cluster $\alpha$ our most general model can be written as
\begin{equation}
	\hat P_e(\vec r) = F(\{ (\vec q_i^{(\alpha)}, \vec v_i^{(\alpha)}) \}_{i \in \alpha};
	                     \{ (\vec q_i^{(\vec r)}, \vec v_i^{(\vec r)}) \}_{|\vec q_i - \vec r|<R};
	                     \mathbf{s}_\alpha, \mathbf{e}_\alpha; \mathbf{a}; \vec r )\,,
\end{equation}
where $\mathbf{s}_\alpha$ are scalar properties describing the cluster,
$\mathbf{e}_\alpha$ are unit vector properties,
$\mathbf{a} \sim \mathcal{N}(\mathbf{0},\mathbb{I})$ is drawn from a standard normal,
and we distinguish between feature tuples and SO(3) vectors using the given notation.
The first argument to $F$ is the set of dark matter particles comprising the cluster, positions and velocities
are evaluated relative to the cluster position and bulk motion respectively.
Conversely, the second argument is the set of particles in the vicinity of the target position $\vec r$,
where the positions are relative to $\vec r$ and the velocities relative to the local bulk motion;
$R$ is a hyperparameter.
Note that operating on particle sets naturally solves the resolution problem mentioned earlier.

DeepSets~\citep{Manzil2017} are a class of architectures that naturally operate on such sets, part of a broader direction in deep learning known as ``geometric deep learning'' (see for example \citealt{bronsteinGeometricDeepLearning2017a,battagliaInteractionNetworksLearning2016,battagliaRelationalInductiveBiases2018,bronsteinGeometricDeepLearning2021} and citations therein).
Geometric deep learning has previously seen success in cosmology in works such as \cite{oladosuMetaLearningOneClassClassification2021,cranmerDiscoveringSymbolicModels2020,cranmerHistogramPoolingOperators2021a,cranmerUnsupervisedResourceAllocation2021}.
Given a tuple of scalars $\mathbf{f}_i$ associated with the $i$th dark matter particle,
a DeepSet first computes another tuple $\mathbf{g}_i$ using a multi-layer perceptron (MLP).
Then a pooling operation (in our case the mean) over the $i$-direction produces a feature tuple 
that is invariant under the ordering of the input particles.
We denote such an architecture as a scalar DeepSet.
We construct the input features $\mathbf{f}_i$ so as to make its elements SO(3) scalars~ \citep{Villar2021}.
This can be achieved by using properties such as $|\vec q_i|$, $|\vec v_i|$, and contractions
between $\vec q_i$, $\vec v_i$ and the elements of $\mathbf{e}_\alpha$.
A simple extension multiplies the $\mathbf{g}_i$ with the $\vec q_i$ before pooling,
thus leading to an output feature tuple in which each element is an SO(3) vector.
We denote such an architecture as a vector DeepSet.
It is easy to see that the described vector DeepSet is rotationally equivariant, since
its output is a linear combination of SO(3) vectors with SO(3) scalar coefficients.
Thus we obtain a rotationally equivariant class of architectures operating directly on the
particle representation instead of gridded fields.

It should be noted that DeepSets exhibit a further feature that turns out to be useful in applications
similar to the one presented here.
Namely, given an appropriate pooling operation, they are to leading order invariant under the number
of input particles.
Thus, training can be performed efficiently on small input sets while inference is then possible,
with typically higher accuracy, on larger sets.
This `sparsification' of input sets during training is also a form of regularization;
in fact, we find it crucial in mitigating overfitting.

\section{Architecture}
\label{sec:arch}

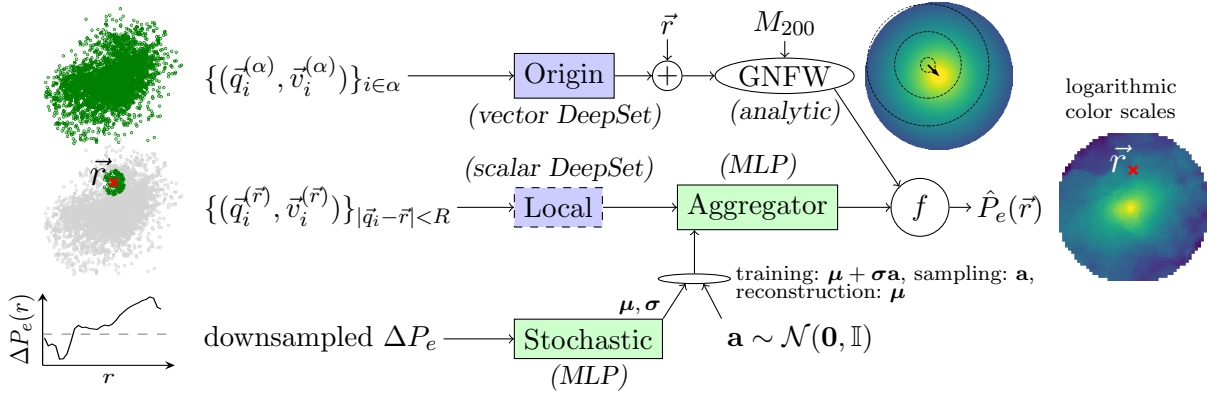
\begin{figure}
\center
\begin{tikzpicture}[xscale=1.2]
\draw (10.5,3.25) node(output) [inner sep=2] {\small $\hat P_e(\vec r)$};
\draw (9.5,3.25) node(combinenode) [circle,draw] {$f$};
\draw [->] (combinenode.east) -- (output.west);
\draw (8,5) node(batt12net) [ellipse,draw,inner sep=1] {GNFW};
\draw (batt12net.south) node [below,yshift=2pt] {\sl\footnotesize (analytic)};
\draw (8,5.7) node(Minput) [inner sep=2] {\small $M_{200}$};
\draw [->] (Minput.south) -- (batt12net.north);
\draw [->] (batt12net.south east) -- (combinenode.north west);
\draw (6.7,5) node(originTNGnode) [circle,draw,inner sep=0] {$+$};
\draw (6.7,5.7) node(TNGposinput) [inner sep=2] {\small $\vec r$};
\draw (5,5) node(originnet) [draw,right,fill=blue,fill opacity=0.2,text opacity=1] {Origin};
\draw (originnet.south) node [below,yshift=2pt] {\sl\footnotesize (vector DeepSet)};
\draw [->] (TNGposinput.south) -- (originTNGnode.north);
\draw (1.5,5) node(haloinput) [right,inner sep=2] {\small $\{ (\vec q_i^{(\alpha)}, \vec v_i^{(\alpha)}) \}_{i \in \alpha}$};
\draw [->] (haloinput.east) -- (originnet.west);
\draw [->] (originnet.east) -- (originTNGnode.west);
\draw [->] (originTNGnode.east) -- (batt12net.west);

\draw (7.7,3.25) node(decodernet) [draw,fill=green,fill opacity=0.2,text opacity=1] {Aggregator};
\draw (decodernet.north) node [above,yshift=-2pt] {\sl\footnotesize (MLP)};
\draw [->] (decodernet.east) -- (combinenode.west);
\draw (5,3.25) node(localnet) [draw,right,fill=blue,fill opacity=0.2,dashed,text opacity=1] {Local};
\draw (localnet.north) node [above,yshift=-2pt] {\sl\footnotesize (scalar DeepSet)};
\draw (1.5,3.25) node(localinput) [right,inner sep=2] {\small $\{ (\vec q_i^{(\vec r)}, \vec v_i^{(\vec r)}) \}_{|\vec q_i - \vec r|<R}$};
\draw [->] (localnet.east) -- (decodernet.west);
\draw [->] (localinput.east) -- (localnet.west);

\draw (5,1.5) node(vaenet) [draw,right,fill=green,fill opacity=0.2,text opacity=1] {Stochastic};
\draw (vaenet.south) node [below,yshift=2pt] {\sl\footnotesize (MLP)};
\draw (1.5,1.5) node(Pthinput) [inner sep=2] [right] {\small downsampled $\Delta P_e$};
\draw (7.3,1.5) node(normalinput) [right,inner sep=2] {\small $\mathbf{a} \sim \mathcal{N}(\mathbf{0},\mathbb{I})$};
\draw [->] (Pthinput.east) -- (vaenet.west);
\draw (vaenet.north east) node(vaenetoutput) [above left,xshift=3pt,yshift=-2pt] {\scriptsize $\boldsymbol{\mu}, \boldsymbol{\sigma}$};
\draw (7,2.3) node(vaecombine) [ellipse,draw,text opacity=0,inner sep=0] {\tiny xxxxx};
\draw [->] (vaenet.north east) -- (vaecombine.210);
\draw [->] (normalinput.north west) -- (vaecombine.330);
\draw [->] (vaecombine.north) -- (vaecombine.north|-decodernet.south);
\draw (vaecombine.east) node [right,text width=4cm,xshift=-2pt,yshift=-2pt,align=left]
	{\lineskiplimit=10pt\lineskip=0pt\scriptsize
         \mbox{training:}~$\boldsymbol{\mu}+\boldsymbol{\sigma}\mathbf{a}$,
         \mbox{sampling:}~$\mathbf{a}$,
         \mbox{reconstruction:}~$\boldsymbol{\mu}$\par};

\draw (haloinput.west) node(haloinputcartoon) [left] {\includegraphics{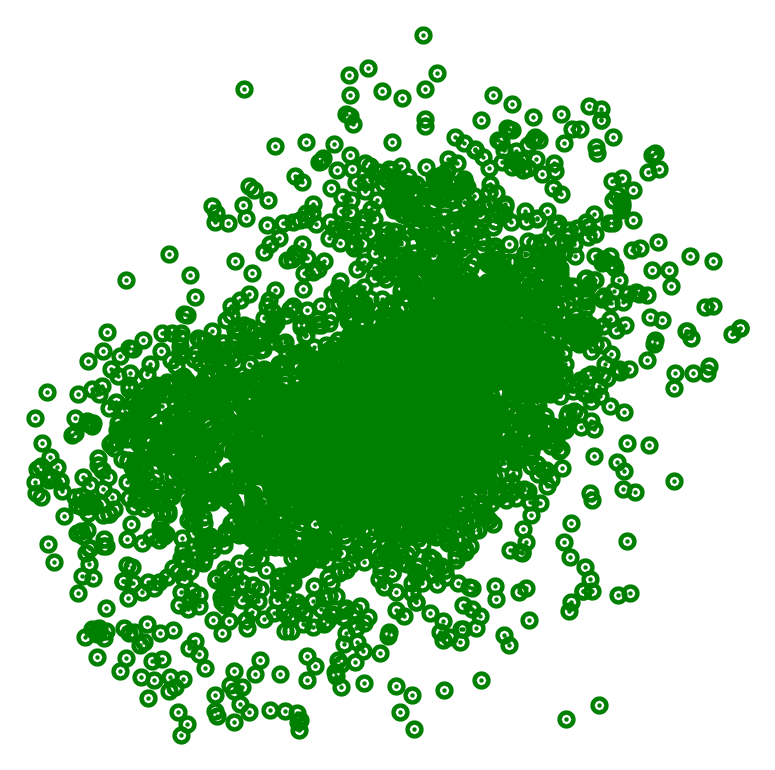}};
\draw (localinput.west) node(localinputcartoon) [left] {\includegraphics{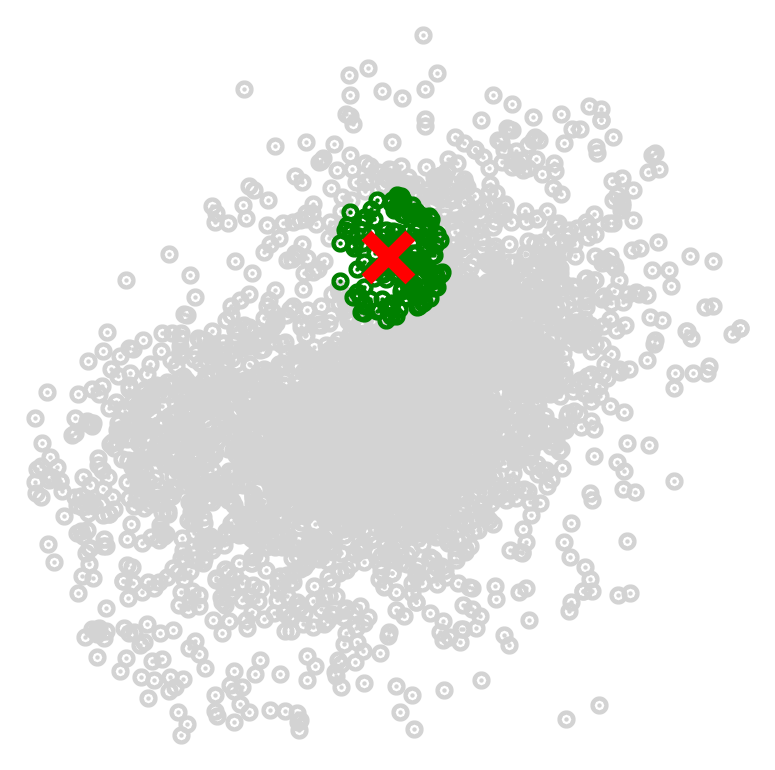}};
\draw (output.east) node(outputcartoon) [right] {\includegraphics{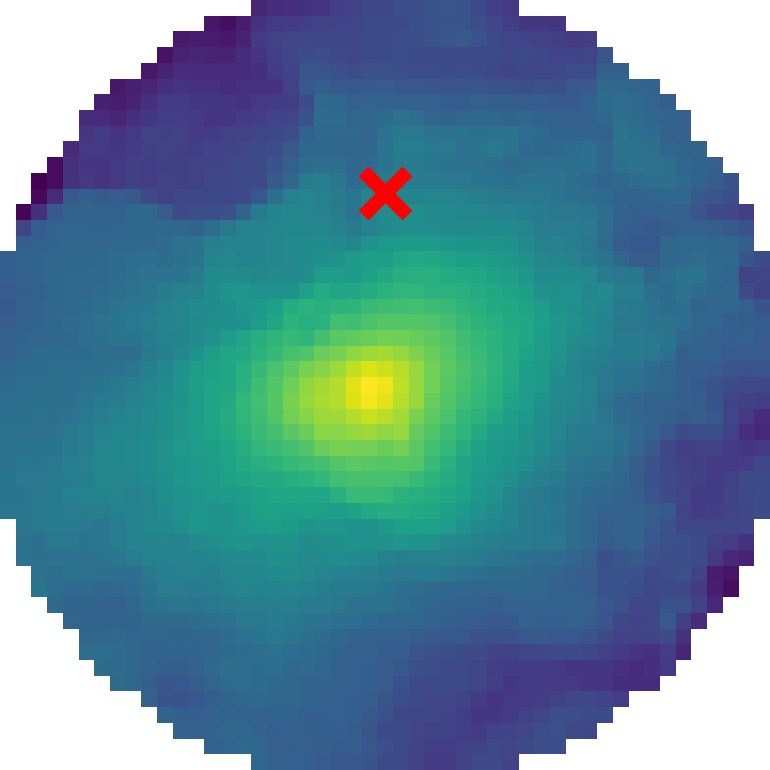}};
\draw (batt12net.east) node(batt12netcartoon) [right] {\includegraphics{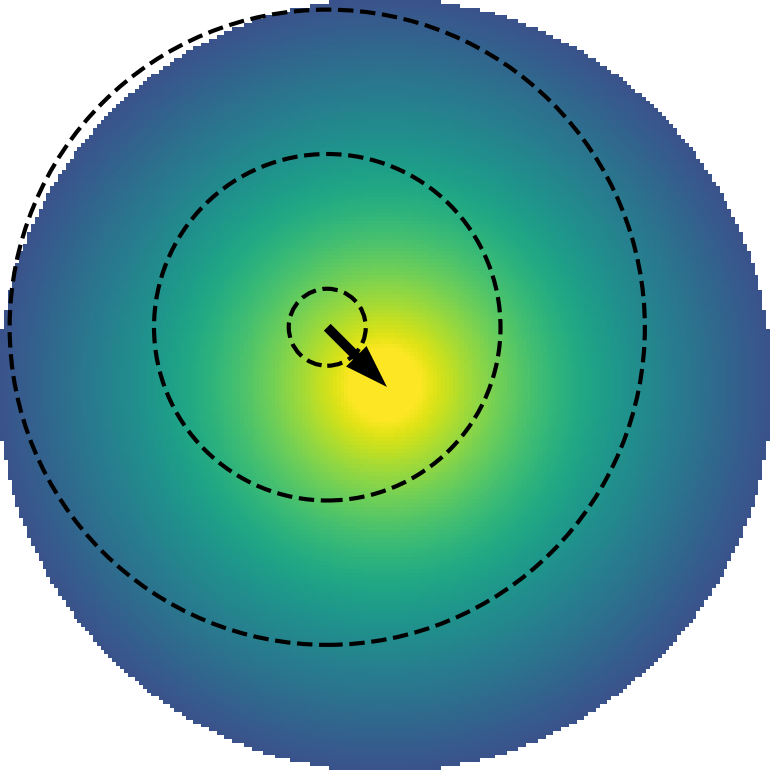}};

\begin{axis}[at={(Pthinput.west)},
             anchor=east,
             xshift=-0.1in,
             yshift=0.05in,
             height=1in,width=1.2in,
             xlabel={\scriptsize \(r\)},
             ylabel={\scriptsize \(\Delta P_e(r)\)},
             axis lines=left,
             ticks=none,
             xmin=0, xmax=2.2,
             ymin=-1.3, ymax=1.7,
             xlabel style={yshift=0.55cm},
             ylabel style={yshift=-1cm}]
	\addplot [color=black, mark=none]
                 table [x=r,y=DeltaPth,col sep=comma] {figures/cartoon_residuals.dat};
        \addplot [color=gray,dashed]
                 coordinates {(0,0)(2.2,0)};
\end{axis}

\draw (11.7,3.9) node [color=white] {\large $\vec r$};
\draw (0.4, 3.73) node [color=black]{\large $\vec r$};


\draw (11,4.7) node(logcomment) [right,text width=1.5cm] {\baselineskip=0pt\scriptsize logarithmic color scales\par};
\end{tikzpicture}
\caption{Schematic illustration of the architecture used in this work.
         Our task is to predict the electron pressure $P_e$ at positions $\vec r$,
	 using phase space coordinates $\vec q$, $\vec v$ of the dark matter particles.
	 The spherically symmetric analytic GNFW profile generally provides a good approximation
	 for the target electron pressure field and also serves as the benchmark in this work.
	 The first correction we introduce is for mis-centering. This correction to the point
	 where the GNFW profile is centered is computed by the \module{Origin} module
	 which is a vector DeepSet operating on all particles in the halo.
	 Second, we introduce local corrections to the GNFW profile using the \module{Local} module
	 which is a scalar DeepSet operating on all particles within a certain radius $R$
	 from the target position.
	 Third, stochasticity is modeled using a conditional-VAE design, in which the \module{Stochastic}
	 module is an MLP acting on a downsampled version of the residuals.
	 The \module{Aggregator} MLP and the function $f$ serve to combine and compress their inputs.
	 Individual modules can be independently removed from and added to the architecture.
	 }
\label{fig:arch}
\end{figure}

Fig.~\ref{fig:arch} schematically illustrates the various architecture components.
We emphasize that most modules can be trained and evaluated independently.
This modular design makes the architecture amenable to interpretation.
At the end of Sec.~\ref{sec:results} we will briefly mention several modules we have
experimentally added to the architecture.
At various points the cluster-scale properties $\mathbf{s}_\alpha$, $\mathbf{e}_\alpha$
are passed, which we omit for conciseness.

The function $f$ produces the final output $\hat P_e(\vec r)$ using two components,
namely a (modified) GNFW prediction and the output of the \module{Aggregator} MLP.
By construction, the GNFW prediction is a single scalar $g$,
while we let the \module{Aggregator} output two scalars, $a_1$, $a_2$.
The function $f$ is chosen as:
\begin{equation}
\hat P_e(\vec r) = f(g, a_1, a_2) = \text{ReLU}[g\,\text{ReLU}(1+a_1) + k \sinh a_2]\,,
\end{equation}
where $k$ is a learnable parameter and $\text{ReLU}(x) \equiv \max(0, x)$ is the rectifying linear unit.
The outer ReLU is chosen because the output is known to be positive, and the inner ReLU because we
believe it unphysical for the GNFW prediction to contribute with inverted sign.
The sinh function provides flexibility in that it can act both as a linear and an exponential,
depending on the hyperparameter $k$, a similar parameterization has already been found
useful in \cite{Thiele2020}.
In some of our experiments we prune the network such that the entire \module{Aggregator} disappears;
in such cases $f$ simply returns the GNFW prediction $g$.

The GNFW model takes as input the target radial position $|\vec r|$, which is corrected for
mis-centering by the \module{Origin} module (the cluster finder estimates cluster positions that are not necessarily
best to center the GNFW profile at).
The \module{Origin} module is a vector DeepSet operating on the input features $|\vec q|$, $|\vec v|$, $\hat q.\mathbf{e}_\alpha$,
$\hat v.\mathbf{e}_\alpha$, and $\mathbf{s}_\alpha$ for particles within $2.5\,R_{200}$.
The $|\vec q|$ and $|\vec v|$ are normalized by $R_{200}$ and $V_{200}$ respectively, and afterwards
transformed to zero mean, unit variance over the training set.
The \module{Origin} module outputs two vectors $\vec o_1, \vec o_2$; the final shift in the
cluster center is then computed as
\begin{equation}
\Delta\vec o = \frac{R_{200}}{2}\tanh[(X_\text{off}/R_{200})\vec o_1 + \vec o_2]\,,
\end{equation}
with $X_\text{off}$ as defined in the following section.
It appears reasonable that in many cases the desired shift in origin will correlate well with
a spatial measure of relaxedness (such as $X_\text{off}$), which motivates the $\vec o_1$ term.
However, in some cases this assumption may be false (e.g., $X_\text{off}$ happens to be very small),
in which case we give the network the freedom to disregard our physical intuition via the separate
$\vec o_2$ vector.
The $\tanh$ acts element-wise and stabilizes training. In the fully trained network, $\Delta\vec o/R_{200}$
is generally small, so the $\tanh$ reduces to a linear (in which case $\Delta\vec o$ transforms as a vector
to a good approximation).
However, due the non-linear effects $\Delta\vec o$ has on other components of the network it is useful
to bound its magnitude to a reasonable range so that
the training process does not become stuck in pathological configurations.

The \module{Aggregator} MLP combines multiple inputs.

The \module{Local} module produces scalar features from the set of dark matter particles in the vicinity of the
target position, where the cutoff $R$ is a hyperparameter.
For each particle we pass the features $|\vec q|$, $|\vec v|$, $\hat q.\mathbf{e}_\alpha$, $\hat v.\mathbf{e}_\alpha$,
and $\vec q.\vec v$, where $\vec q$ and $\vec v$ are with respect to the target position
and local bulk velocity respectively
and their magnitudes normalized to zero mean, unit variance over the training set.
After passing the local particles through the DeepSet, we concatenate the resulting tuple with the
number of particles within $R$ to set the scale.
If other cluster-scale information, besides the \module{Local} module, is passed to $f$, we also pass information
on the target position $\vec r$ to the \module{Aggregator}.
This is achieved by passing the scalars $|\vec r|$ and $|\hat r.\mathbf{e}_\alpha|$, with $|\vec r|$ normalized by $R_{200}$,
as well as the $\mathbf{s}_\alpha$.
The use of the \module{Local} module is a form of expanding our training set
(one could imagine compressing the set of all particles comprising the cluster into some code which is then
evaluated at different positions; in this case, the training set would be very small however).

The other input to the \module{Aggregator} models the probabilistic nature of the mapping,
through a conditional VAE~\citep{Johnson2016,Esser2018,Lanusse2021,Horowitz2021} architecture.
The \module{Stochastic} module is the standard VAE encoder, taking as input the residuals of the electron pressure field
with respect to a deterministic model.
Since the simulation we are using implements AGN feedback in a spherically symmetric fashion,
we average these residuals in spherical shells around the cluster position (another instance of domain knowledge).
These spherical shells are chosen with respect to the position computed by the Rockstar halo finder
(see the following section), i.e., the \module{Origin} module has no influence on how these shells are chosen.
These shells are chosen as 32 linearly spaced radial intervals out to $2R_{200}$.
We perform principal component analysis on the residuals over the training set and find that the most important
eigenvector represents transfer of electron pressure between inner and outer parts of the cluster,
bolstering our intuition that the stochasticity is driven by AGN feedback.
We will find in Sec.~\ref{sec:results} that this interpretation is perhaps too simplistic, however.
Given our expectation that the unresolved AGN activity drives stochasticity,
we choose the VAE code to be a single number; such a small latent space dimension is also useful in
regularizing the model in view of the small training set.
We have explicitly confirmed that doubling the latent space dimension leads to worse results.

On a fundamental level, our architecture composes many MLPs.
While we allow for some degree of heterogeneity in their structure,
we generally use LeakyReLU activation functions, layer normalization~\citep{Ba2016},
depths of $3 \cdots 4$ hidden layers, and $\sim 196$ hidden neurons per layer.
We find that, not surprisingly, the \module{Aggregator} is most susceptible to overfitting.
This motivates us to introduce dropout~\citep{Hinton2012} there,
with rates of $\sim 5\,\%$ found optimal in the hidden layers
while the first layer requires much higher rates of $\sim 30\,\%$.

\section{Data and training}
\label{sec:methods}

We use the IllustrisTNG 300-1
simulation~\citep{IllustrisTNG1,IllustrisTNG2,IllustrisTNG3,IllustrisTNG4,IllustrisTNG5,Nelson2019}
for training and testing.
This simulation provides a gravity-only and a full-physics run with the same initial conditions.
In this work, we restrict ourselves to the present-day (redshift $z=0$) snapshot;
a generalization to earlier times is naturally possible by passing $z$ to the various MLPs in the architecture.

We use the state-of-the-art cluster finder code Rockstar~\citep{Behroozi2012,Behroozi2013}
to identify clusters with masses $M_{200} > 5 \times 10^{13}\,M_\odot/h$
in the gravity-only snapshot.\footnote{The reason
for this choice of mass cutoff is that in the IllustrisTNG astrophysics model
at lower masses the gas physics changes qualitatively as AGN feedback is more
effective in driving gas out of the cluster.}
The resulting 463 clusters are randomly assigned to training (70\,\%),
validation (20\,\%), and testing (10\,\%) sets.
They have radii $R_{200}$ ranging from $600$ to $1600\,\text{kpc}/h$
and contain between $1.5$ and $47$ million dark matter particles within $2.5 R_{200}$.

The cluster-scale scalars $\mathbf{s}_\alpha$ are chosen as:
\begin{itemize}
\item logarithmic mass, $\log M_{200}$,
\item position offset from center of mass, $X_\text{off}$,
\item velocity offset from bulk motion, $V_\text{off}$,
\item same as $X_\text{off}$, but with respect to the center of mass of \emph{all} particles
      within $2.5\,R_{200}$ (instead of only the ones identified by Rockstar as being bound),
\item magnitude of angular momentum, $|\vec J|$,
\item eigenvalues of mass inertia tensor, $I^{(m)}_a\,(a=1,2,3)$,
\item eigenvalues of velocity inertia tensor, $I^{(v)}_a\,(a=1,2,3)$
      (sometimes called the velocity dispersion tensor).
\end{itemize}
The cluster-scale vectors $\mathbf{e}_\alpha$ are chosen as:
\begin{itemize}
\item angular momentum, $\vec J$,
\item position offset, $\vec X_\text{off}$,
\item eigenvectors of mass inertia tensor, $\vec I^{(m)}_a\,(a=1,2,3)$,
\item eigenvectors of velocity inertia tensor, $\vec I^{(v)}_a\,(a=1,2,3)$.
\end{itemize}
It is important to ensure the correct behaviour of these quantities under SO(3) transformations.
Thus, we order eigenvalues, together with the corresponding eigenvectors, according to magnitude.
The orientation of the eigenvectors needs also to be fixed.
This can be accomplished using any pseudo-vector, in this case we choose orientation such that
the contraction of any eigenvector with the angular momentum $\vec J$ is positive.

All vectorial properties $\mathbf{e}_\alpha$ are normalized such that contractions with isotropically
distributed unit vectors have unit variance.
The scalar properties, except for $\log M_{200}$, are normalized by the appropriate self-similar
scales; using $M_{200}$ and $R_{200}$, any quantity can be nondimensionalized.
Thus, we reduce all inputs to geometric ones apart from $\log M_{200}$ which sets the scale.
After the self-similar normalization, we apply the affine transformation to the $\mathbf{s}_\alpha$
that yields zero mean and unit variance over the training set.

Since our training set is relatively small, we find it crucial to add noise to the cluster-scale
properties.
The non-uniform distributions of some of the scalars $\mathbf{s}_\alpha$ motivate us to use the
following procedure.
For each feature in $\mathbf{s}$, we sort the values in the training set and compute the
nearest-neighbour differences $\Delta_+, \Delta_-$.
During training, we generate noise according to
\begin{equation}
\delta = N r \Delta_{\text{sgn}(r)}\,,\,r\sim\mathcal{N}(0,1)\,.
\end{equation}
Here, $N$ is a hyperparameter for which we find values $\mathcal{O}(10)$ to work best.
Owing to its exceptional role as setting the overall scale, we allow for a different value of $N$
for the $\log M_{200}$ feature, finding that somewhat lower noise levels are optimal.
Note that this noise prescription does not actually preserve the mean, but empirically we find
it to work better than `patching' two one-sided Gaussians.
For the vector properties, we experimented with rotations by angles drawn from a normal.
We find mild evidence that standard deviations of $\sim2\,\text{deg}$ are preferred over noiseless
vectors.

We produce electron pressure fields from the full-physics simulation using
Voxelize~\citep{Paco2021CMD}, with a voxel sidelength of $5 R_{200} / 64$.
This resolution is sufficient for current and near-future tSZ measurements;
we find that higher resolutions lead to more unstable training, presumably due to large local outliers.
The choice to keep the resolution in units of $R_{200}$ is motivated by the general
geometric spirit of this work, a production-quality model would most likely choose a constant
resolution instead.

Our reconstruction loss for a given cluster is given by
\begin{equation}
\mathcal{L}_\text{recon} = \left\langle \left(
                           \frac{P_e(\vec r)-\hat P_e(\vec r)}{P_{200}}
			   \right)^{\!\!2} \right\rangle_{|\vec r| < 2R_{200}}\,,
\label{eq:Lrecon}
\end{equation}
where the normalization with $P_{200}$ mitigates our dearth of clusters
at the high-mass end.\footnote{For practical applications the scaling with $P_{200}$
should be omitted, in which case a somewhat larger training set will likely be required.}
The target positions $\vec r$ are randomly sampled during training for efficiency,
while testing is of course performed on all available voxels.

Training is performed using the Adam optimizer~\citep{Kingma2015}
and a one-cycle learning rate schedule~\citep{Smith2017}.
Each epoch iterates once through all clusters in the training set,
randomly choosing 256 target positions in each cluster and assembling 4 clusters into each mini-batch.
We typically train for 100 epochs, in some cases 200.
We find it necessary to choose different learning rates for the individual modules.
For modules that have access to cluster-scale information (\module{Aggregator}, \module{Origin}),
we also apply weight decay.
We find it beneficial to apply gradient clipping (typically at $\mathcal{O}(1)$) in order to stabilize training.

It turns out to be optimal to pass $\mathcal{O}(10^3)$ randomly sampled dark matter particles
to the \module{Origin} module during training, since such a sparse sampling reduces overfitting.
We typically pass $\mathcal{O}(10^2)$ dark matter particles to the \module{Local} module,
this is purely for efficiency.
In many cases there are fewer dark matter particles than this number in the local vicinity;
then we simply sample some particles multiple times.
During testing we increase these numbers typically by a factor $\mathcal{O}(10)$ and observe
some benefit from the finer sampling.
Further increase of the number of dark matter particles passed does not increase the prediction
quality.

For hyperparameter searches we use the Optuna package~\citep{Akiba2019}, solving the problem
\begin{equation}
	\theta_\text{opt} = \text{argmin} \mathcal{L}_\text{opt}(\theta)
	\ \text{with} \
	\mathcal{L}_\text{opt}(\theta) \equiv \text{median}(\mathcal{L}_\text{recon}[\text{network}_\theta]
	                                                    / \mathcal{L}_\text{recon}[\text{GNFW benchmark}])\,,
	\label{eq:lossquantifier}
\end{equation}
where the median is over the validation set at the end of training.
During training runs on architectures in which the stochastic module is included, we take as the training
loss the sum of reconstruction loss and negative KL divergence of the VAE code with respect to a standard normal,
the latter multiplied with a scaling which we anneal from zero to some hyperparameter over the course
of training.
For such architectures, we perform multi-objective optimization on both $\mathcal{L}_\text{opt}$
and the mean of the KL divergence over the validation set.
We then produce figures similar to Fig.~\ref{fig:scatterloss} for multiple models in the Pareto frontier,
considering the validation set only, in order to choose the best model.\footnote{Total compute cost
is 13.4~(Tesla P100+9CPU)~khr (1.09t $\text{CO}_2$e~\citep{Lannelongue2021})
with a PyTorch~\citep{PyTorch} implementation.}

\section{Results and Discussion}
\label{sec:results}

\begin{figure}
	\center
	\includegraphics[width=0.97\textwidth]{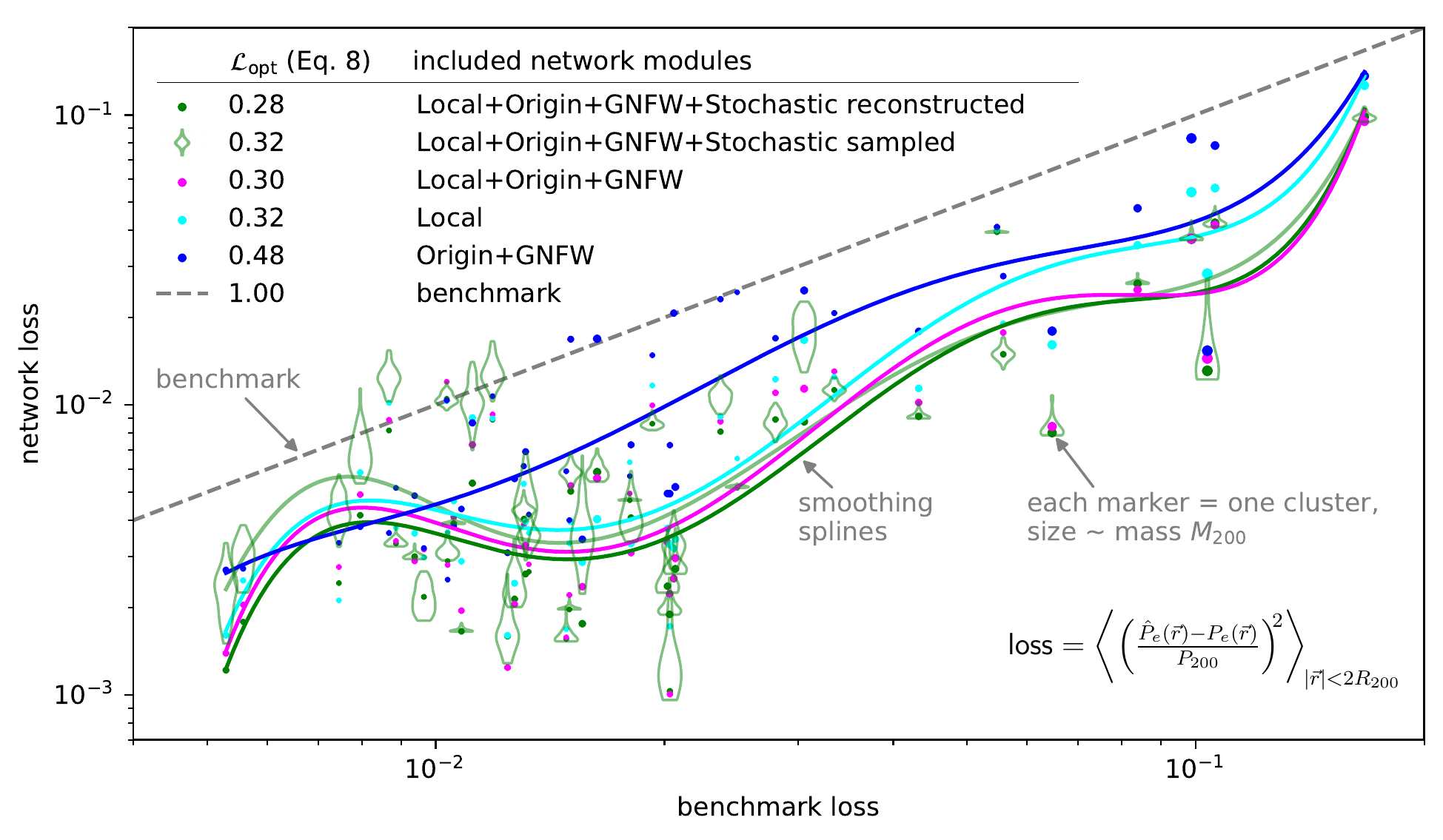}
	\caption{Network losses evaluated on testing set and compared against the GNFW benchmark model.
	         Each data point is an individual cluster, the marker size indicating mass.
		 For the version of the network including the \module{Stochastic} module,
		 two modes of evaluation are possible:
		 First, we can evaluate the reconstruction, i.e., the \module{Aggregator} receives
		 the output $\boldsymbol{\mu}$ of the \module{Stochastic} MLP (dark green points);
		 second, we can randomly sample the VAE code, i.e., the \module{Aggregator} receives
		 the vector $\mathbf{a}$ drawn from a standard normal (light green violins).
		 The lines are simple smoothing splines and only meant to guide the eye
		 (the light green line corresponds to the means of the violins).
		 The numbers in the legend's first column are the performance metric $\mathcal{L}_\text{opt}$
		 introduced in Eq.~\ref{eq:lossquantifier} (lower is better, $\text{benchmark} \equiv 1$).}
	\label{fig:scatterloss}
\end{figure}

In Fig.~\ref{fig:scatterloss} we plot several network losses compared against the GNFW benchmark.
Only correcting for mis-centering (blue) already gives a factor $\sim 2$ improvement over the use of cluster centers
as identified by Rockstar.
Likewise, only using the dark matter matter particles in the vicinity of the evaluation point\footnote{We find that
$R \sim 300\,\text{kpc}/h$ is a good choice.} (cyan) yields a further improvement.
Combining the local information with the shifted GNFW profiles (magenta) performs better than \module{Local}-only by a few percent,
the improvement being most pronounced in the high-loss regime.
We conjecture that this could be because the addition of the simpler GNFW model helps the network generalize in
these relatively rare situations.
Expectedly, the model including the \module{Stochastic} module (green) generally obtains lower reconstruction losses than the other models.
The corresponding losses with random VAE samples (light green) are not much worse in most cases,
although a larger training set would certainly help the network learn
a more robust representation of the probabilistic component.

Naturally, we should ask whether our models are learning something trivial.
We have checked that a more general spherically symmetric model,
implemented as an MLP that takes as input $|\vec r|$ and the cluster
scalars $\mathbf{s}_\alpha$, does not perform more than a few percent better than the GNFW benchmark.
Similarly, we find that a network using only the local density achieves
more than twice the loss $\mathcal{L}_\text{opt}$
compared to the \module{Local} network, demonstrating that the DeepSet is providing substantial information.

We have also experimented with adding further modules to the network.
First, between \module{Origin} and GNFW we have inserted an MLP
that uses the cluster $\mathbf{s}_\alpha$, $\mathbf{e}_\alpha$
to account for deviations from spherical symmetry.
We find no improvement from this modification.
Second, we have constructed vector and scalar DeepSets
operating on the cluster set $\{ (\vec q_i^{(\alpha)}, \vec v_i^{(\alpha)}) \}_{i \in \alpha}$
whose outputs were then passed to the \module{Aggregator}.
Since these additional modules also do not yield any improvements,
we conclude that the relatively large local regions contain enough information
to infer the global properties of the cluster.
It is important to appreciate that even these null results can tell us something physical,
again a consequence of the interpretable, modular design.

\begin{figure}
\begin{minipage}[b]{0.5\textwidth}
\begin{tikzpicture}
	\node (images) at (0, 0) {\includegraphics[width=\textwidth]{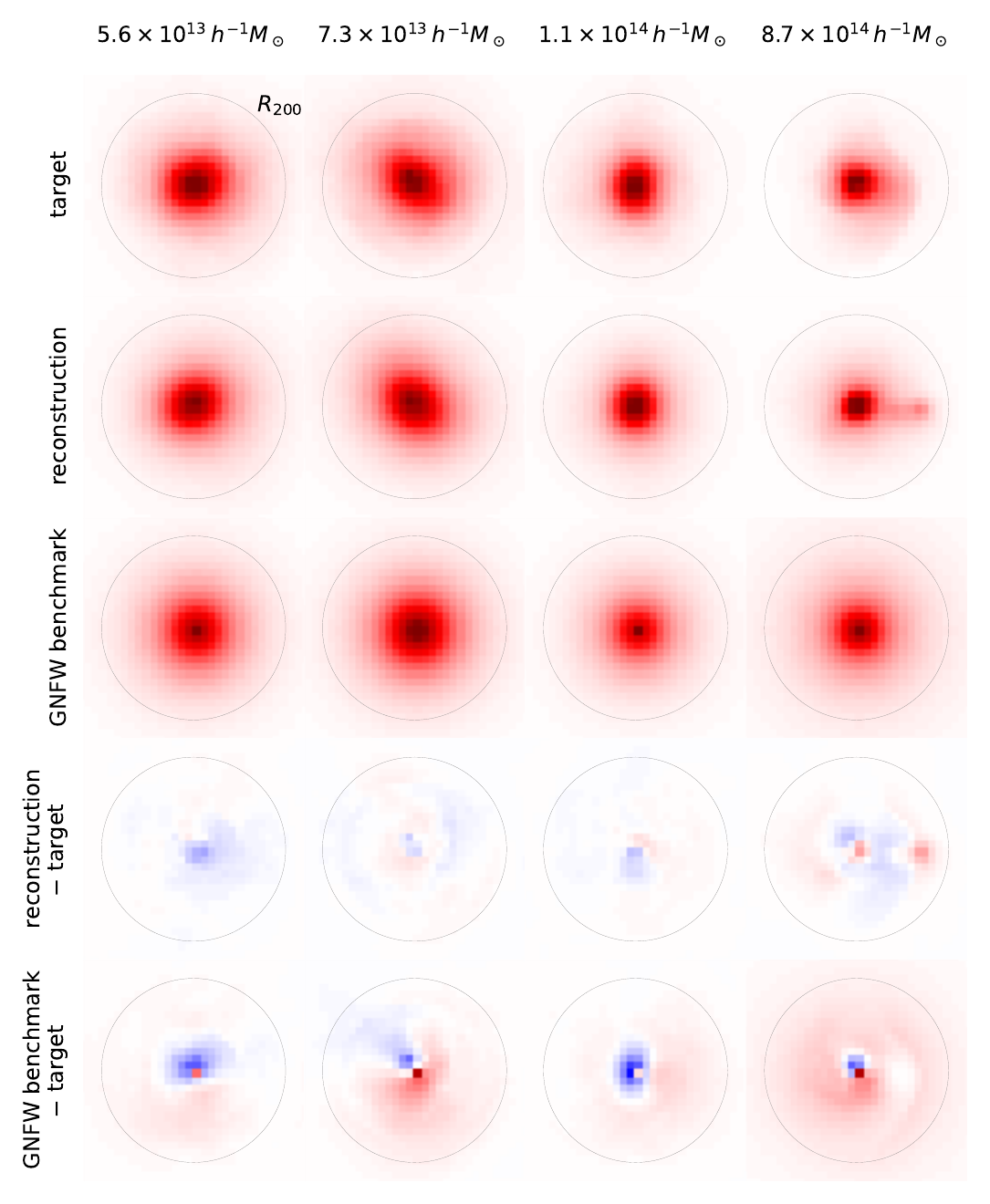}};
	\node (cbar) at (images.north east) [right,xshift=-0.5cm,yshift=-0.75cm] {\includegraphics[width=0.6\textwidth]{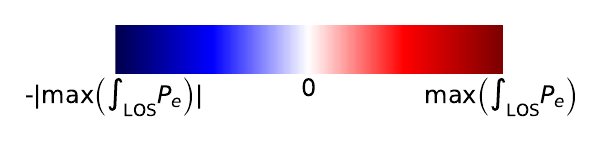}};
\end{tikzpicture}
\end{minipage}%
\begin{minipage}[b]{0.5\textwidth}
	\captionof{figure}{Projected electron pressure fields of four clusters.
	                   The projection directions were chosen randomly.
			   Each `circle' has radius $2 R_{200}$.
			   Cluster mass increases from left to right.
			   The color scale is consistent within each column (i.e., for each cluster),
			   normalized to the maximum pixel.
			   The agreement between the target and the neural network
			   prediction is good and always better than the GNFW benchmark's performance.
			   Note the dipole pattern in the GNFW residuals in the second and fourth
			   column, indicating the effect of mis-centering.
			   Deviations from spherical symmetry (compare the GNFW prediction)
			   are generally predicted in qualitative agreement with the target.
			   However, the predicted field is generally too smooth, which is
			   a common problem with neural networks.
			   In Appendix~\ref{sec:compareCNN}, we perform a qualitative comparison
			   with results obtained using the CNN from \cite{Thiele2020}.}
	\label{fig:images}
\end{minipage}
\end{figure}

\begin{figure}
\begin{minipage}[b]{0.5\textwidth}
	\includegraphics[width=\textwidth]{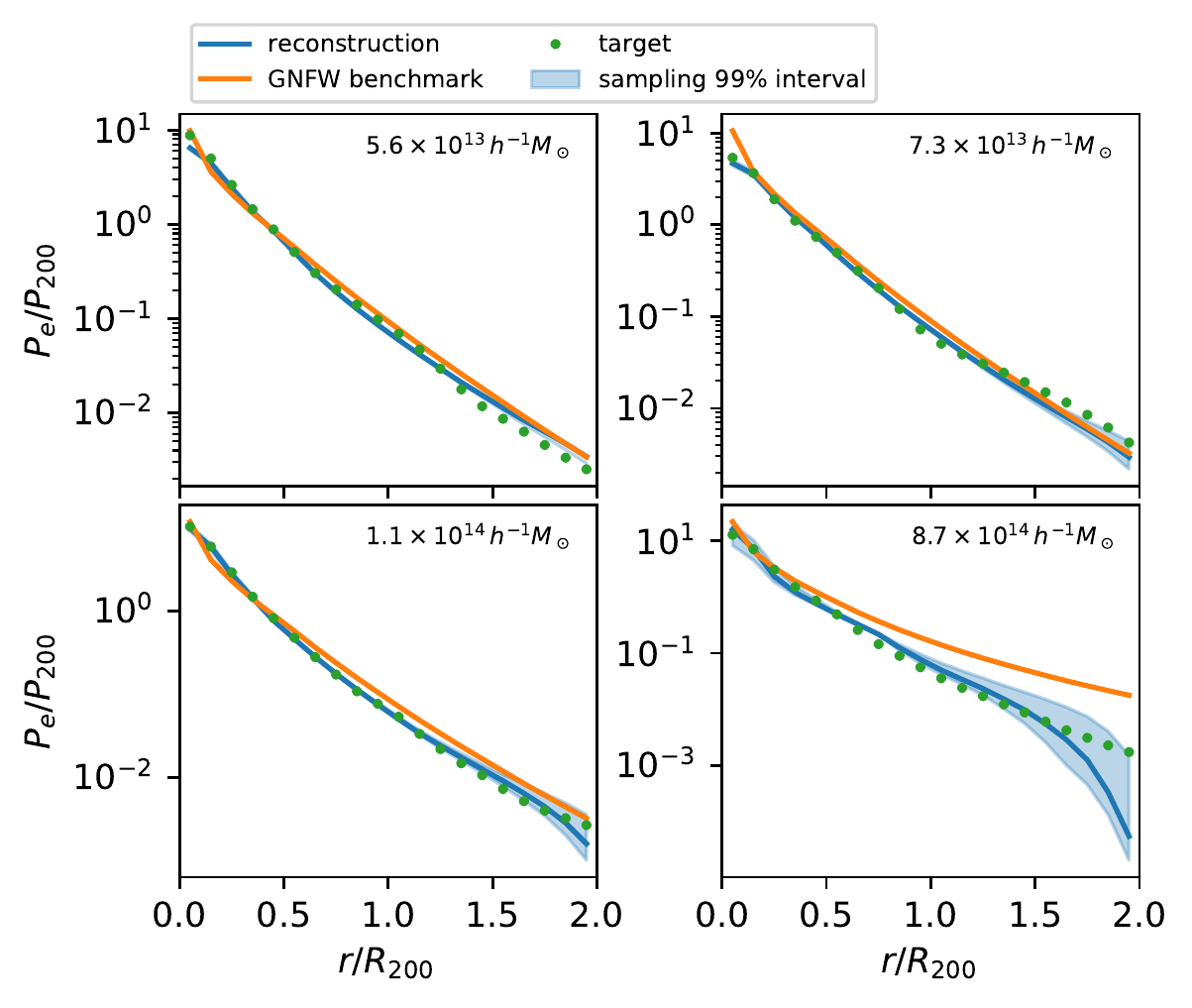}
\end{minipage}%
\begin{minipage}[b]{0.5\textwidth}
	\captionof{figure}{Electron pressure profiles of the same clusters as displayed
	                   in Fig.~\ref{fig:images}.
			   The network predictions are generally in better agreement with
			   the target than the GNFW profiles.
			   For all but the least massive objects the network appears to have
			   learned a good estimate of its uncertainty (represented by the blue bands),
			   with the predicted uncertainty increasing towards the cluster outskirts
			   where substructure and the two-halo term start to become important.
			   The neural network is able to predict a large dynamic range of electron
			   pressure values, in particular the rare, dense cluster cores are
			   predicted accurately.}
	\label{fig:profiles}
\end{minipage}
\end{figure}

In order to gain some further intuition, we now present two plots focusing on four example clusters from
the testing set.
These clusters were chosen at percentiles 25, 50, 75, and 100 in $M_{200}$ to give a representative sample.
We use the most powerful network for the predictions, corresponding to the green markers and violins
in Fig.~\ref{fig:scatterloss}.

First, in Fig.~\ref{fig:images} we show images of projected electron pressure.
Note that the cutoff $r < 2 R_{200}$ has been applied in all rows, so the projection depth varies with
distance from the cluster center.
Visually, the spherically symmetric GNFW benchmark model provides a reasonable approximation to the target
for all but the most massive object.
The network predictions do generally pick up deviations from spherical symmetry in approximately the right
direction.
However, the network predicts fields that are noticeably smoother than the target;
this is a commonly observed problem in similar tasks~\citep{Rothschild2021}.
The most massive object (right column) appears to have undergone a recent merger event.
Our network does realize this and provides a relatively good prediction, although the detailed structure
of the electron pressure field around the secondary halo is not perfect.
On the other hand, the GNFW benchmark struggles with this sample and overpredicts the pressure almost everywhere.

Second, in Fig.~\ref{fig:profiles} we plot the electron pressure profiles corresponding to the same objects.
This figure may be somewhat easier to interpret than Fig.~\ref{fig:images}, and it also gives us the opportunity
to explore the probabilistic component of our network.
We observe that generally the profiles predicted by our network are in better agreement with the target than
the GNFW benchmark.
It is noticeable that the GNFW profiles are generally too high around $r \sim R_{200}$,
although we have carefully fitted them to the same data the network was trained on.
This is a consequence of the fact that averaging in spherical shells and the locally evaluated loss function
Eq.~\ref{eq:Lrecon} do not commute in general.
The stochastically sampled network predictions include the reconstructed profile in all four examples;
this is a good indication that the network's representation of stochasticity is in fact a reasonable one.
Furthermore, our expectation that the stochastic component is mainly responsible for a transfer of pressure
between inner and outer parts of the cluster is confirmed.
However, it should be noted that the predicted stochasticity in the profiles increases with cluster mass.
This is difficult to reconcile with our argument that the stochasticity is primarily driven by the unresolved
AGN activity, since AGN feedback is more efficient in altering the gas distribution in the shallower gravitational
potential wells of lower-mass objects.
This may be an indication that the primary driver of stochasticity is in fact the mass accretion rate,
as was argued in \cite{Rothschild2021}.
Such an interpretation is further supported by the fact that the merger in the rightmost column of
Fig.~\ref{fig:images} seems to be the most challenging scenario for the network.

\begin{figure}
\begin{minipage}[b]{0.6\textwidth}
	\includegraphics[width=\textwidth]{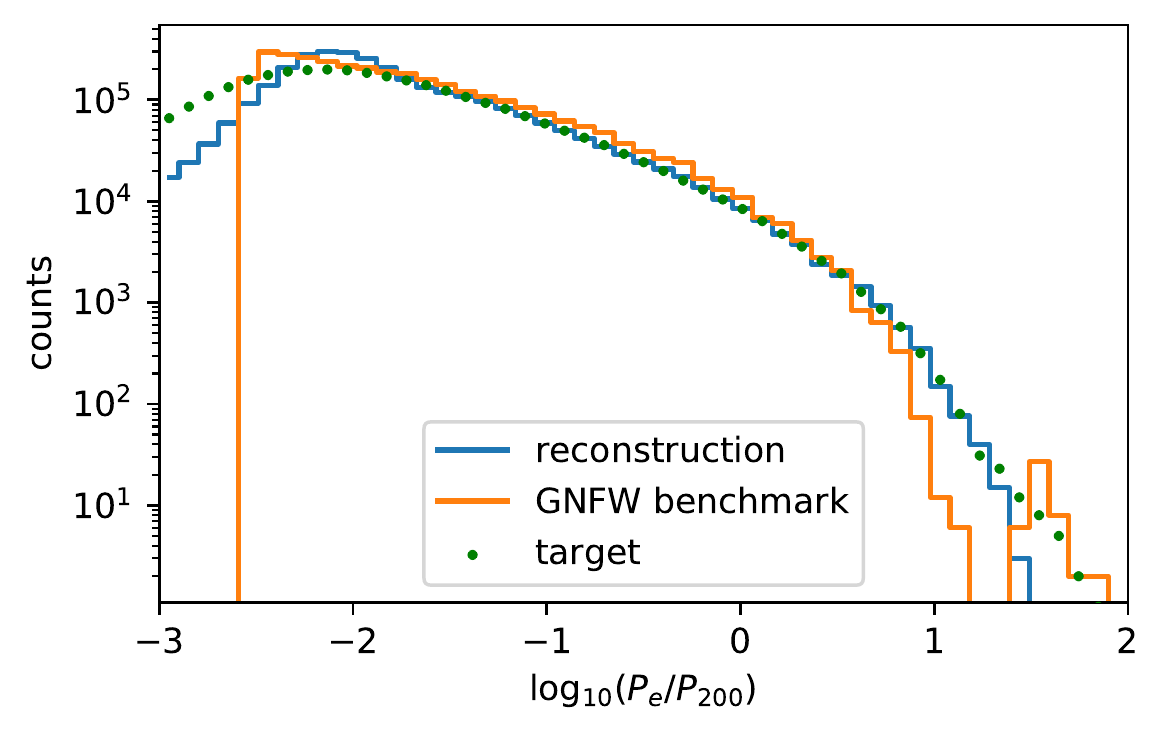}
\end{minipage}%
\begin{minipage}[b]{0.4\textwidth}
	\captionof{figure}
	        {Histograms of electron pressure. We directly histogram the voxelized electron pressure
	         values (c.f. Sec.~\ref{sec:methods}). Thus, this figure is for illustrative purposes only,
		 since the voxel sizes vary between different clusters (as they are fixed in units of $R_{200}$).
		 Regardless, this plot is a useful diagnostic.
		 The neural network is able to predict a large dynamic range of electron pressure values,
		 performing much better than the GNFW benchmark.
		 Low values are underpredicted because they contribute little to the loss function,
		 while high values are overpredicted presumably due to their extreme sparsity.}
	\label{fig:pdf}
\end{minipage}
\end{figure}

As a final result, we examine the histograms of electron pressure values, shown in Fig.~\ref{fig:pdf}.
We observe that for the range $-1.8 \lesssim \log_{10}(P_e/P_{200}) \lesssim 1$ the network produces
a histogram that is in almost perfect agreement with the target, while the GNFW benchmark shows some
discrepancies.
At low electron pressure, the network predictions cluster in a relatively narrow range
$-2.3 \lesssim \log_{10}(P_e/P_{200}) \lesssim -1.8$ below which the histogram drops off compared to the target.
This behavior is explained by the small contribution such small electron pressures give to the loss function.
Similarly, for $\log_{10}(P_e/P_{200}) \gtrsim 1.2$, the network shows a lack of high-pressure values.
This is most likely because predicting this small number of extreme peaks is a difficult objective given
the small training set.

\section{Future directions}
\label{sec:future}

We have developed a general method to construct interpretable models that predict continuous
fields from a set of points while enforcing the underlying symmetries.
We have argued that the set-based approach introduced in this work has multiple advantages over
CNNs that are commonly chosen for similar problems.
Specifically, the direct operation on the simulation representation as a set of dark matter particles simplifies
\begin{itemize}
	\item incorporation of inductive biases
	      (such as the prominent role clusters play for the electron pressure field),
	\item modularization of the architecture into physically meaningful units,
	\item enforcement of underlying symmetries,
	\item fast training, through `sparsificiation' of training data.
\end{itemize}

The application to cosmological structures demonstrates the power of our approach and opens up
several directions of further investigation.
The \module{Origin} module could possibly be almost directly incorporated in existing codes to improve
the centering of spherically symmetric profiles.
In order to get this work to a production-quality stage, training would need to be extended to lower-mass
halos and non-zero redshifts, both of which should not present any major obstacles.
Furthermore, it may be useful to combine the trained network with a CNN that handles the low-density
regions outside the most massive objects.
Such a combined architecture could then be used to predict cosmologically interesting summary statistics.
Extensions to other baryonic fields beyond electron pressure should be straightforward and could
potentially be easier since electron pressure is comparatively sparse.
For fields that are better correlated with the dark matter density than is electron pressure,
CNNs may benefit from the high locality. However, this would also apply to our set-based architecture,
in which the cluster-scale modules could potentially be discarded in such cases.
An interesting extension could involve transfer learning to sub-grid implementations different
from IllustrisTNG.

In terms of interpretation, symbolic regression of e.g., the \module{Origin} module could provide useful insights
and perhaps compact equations that would be more convenient to implement in existing codes.
A more detailed ablation study, e.g., on the importance of the various cluster-scale properties $\mathbf{s}_\alpha$,
could be interesting.

Beyond cosmology, we see potential use cases in irregular structures in condensed matter
or in super-resolution atmosphere models from scattered meteorological measurements.

\acknowledgments
This work made use of the following software packages:
\texttt{numpy} \citep{numpy},
\texttt{scipy} \citep{scipy},
\texttt{jupyter} \citep{jupyter},
\texttt{sklearn} \citep{sklearn},
\texttt{matplotlib} \citep{matplotlib},
\texttt{torch} \citep{PyTorch},
\texttt{torch\_geometric} \citep{torch_geometric},
and
\texttt{astropy} \citep{astropy}.

\bibliography{main}{}
\bibliographystyle{aasjournal}

\appendix

\section{Qualitative comparison to CNN}
\label{sec:compareCNN}

In this section, we provide a qualitative comparison to the convolutional neural net (CNN)
results from \cite{Thiele2020}.
As we have argued in the main text, the set-based architecture is better able to structurally
mirror the distinct features of the problem, namely sparseness and rotational equivariance.

Showing that these advantages also translate into better predictions compared to those obtained
with CNNs in \cite{Thiele2020} is a non-trivial task, however.
In that earlier work, the emphasis was on predicting an entire volume, while here we are
constraining ourselves to massive halos (we have argued, however, that the remainder of the task
should be relatively straightforward and can probably be solved with CNNs).
Thus, \cite{Thiele2020} optimized the network for summary statistics (mostly the power spectrum)
and worked at a fixed spatial resolution (while we, in this work, choose to adapt the resolution
to individual halos, in the geometric spirit of the architecture).

\begin{figure}
\includegraphics[width=\textwidth]{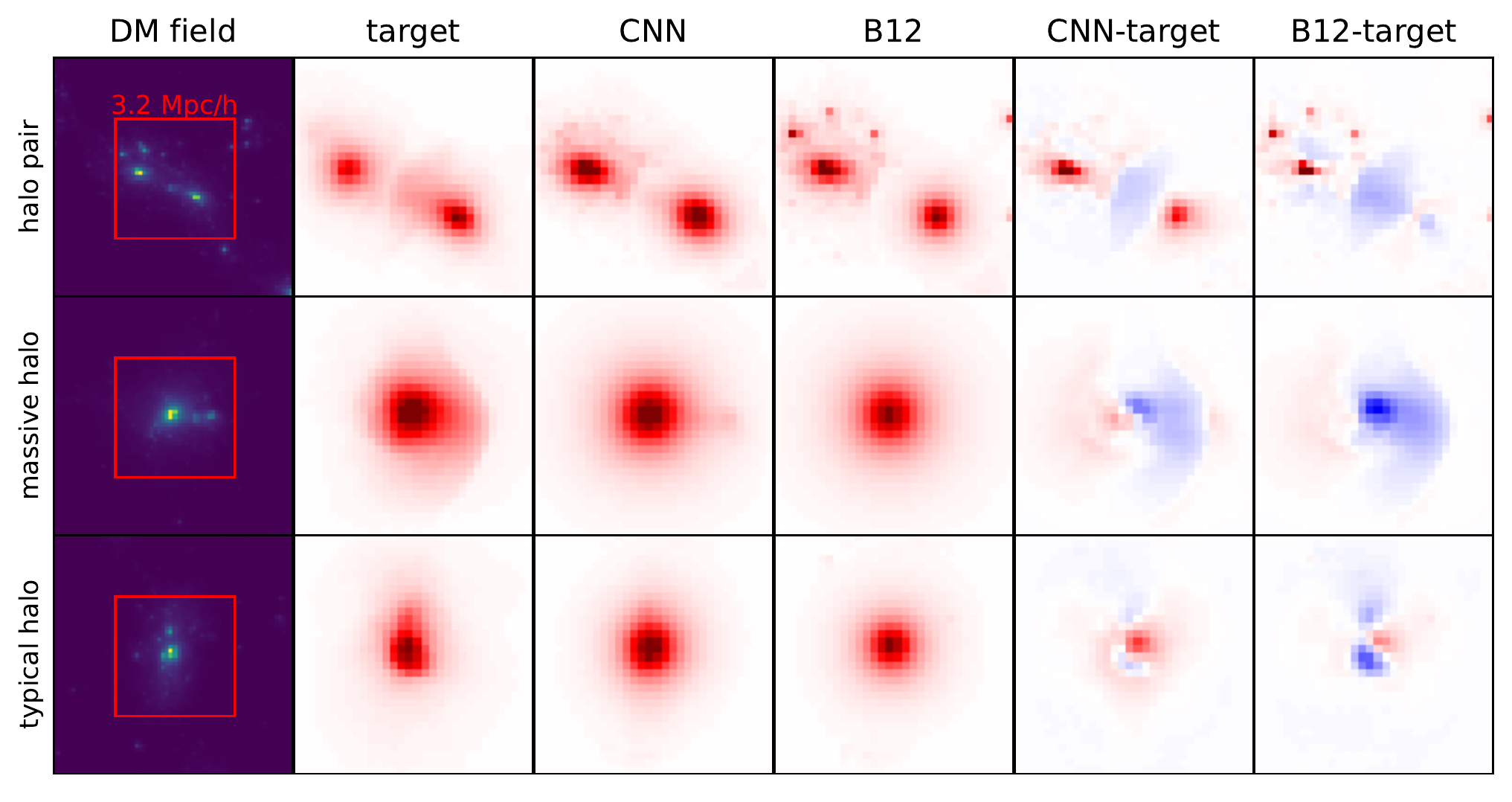}
\caption{For comparison with Fig.~\ref{fig:images},
         we provide images of some halos obtained with the CNN from \cite{Thiele2020}.
         See the text for cautionary notes on direct comparison.}
\label{fig:CNN_images}
\end{figure}

Due to these caveats, a comprehensive quantitative comparison is beyond the scope of this work
(and should probably be performed at the level of summary statistics one would eventually be interested
in, which necessitates inclusion of redshift dependence).
However, a relatively straightforward comparison can be made at the level of halo images.
We have already presented integrated images of a few halos, obtained with our set-based network,
in Fig.~\ref{fig:images}.
For comparison, Fig.~\ref{fig:CNN_images} provides similar images obtained with the CNN
(this figure is a modified version of Fig.~10 in \cite{Thiele2020}).
The color scales are constructed identically to the way chosen in Fig.~\ref{fig:images}.
We need to mention a few points in which the two figures are not directly comparable.
First, the CNN images are at a fixed resolution, with $3.2\,h^{-1}\text{Mpc}$ sidelength.
Second, the semi-analytic model labeled `B12' is similar to the GNFW model used as a benchmark in this work,
but it was not refit to the specific simulation (which is IllustrisTNG-300, same as in this work).
Only an overall scaling was applied to the `B12' pressure profiles.

Despite these differences, at least a qualitative comparison is possible.
The object labeled `massive halo' is actually the same object as the right-most one in Fig.~\ref{fig:images},
with the same projection direction (it is a lucky coincidence that this object ended up in the testing
set in \cite{Thiele2020}, too).
Here, we observe that the CNN tends to make the same mistakes as the B12 model, albeit to a lesser extent.
This strong reliance on the semi-analytic model is a general trend observed with the CNN,
while the set-based architecture proves to be relatively robust to mistakes made by the GNFW model.
At a more quantitative level, the set-based architecture presented in this work produces a more accurate
prediction than the CNN, as evidenced by the darker residuals in Fig.~\ref{fig:CNN_images} compared to
Fig.~\ref{fig:images}.

While for the other two objects in \ref{fig:CNN_images} we unfortunately do not have counterparts
in Fig.~\ref{fig:images}, the same trends appear to hold: the CNN is overly reliant on the B12 model,
which translates into larger errors than those made by the set-based architecture.

\end{document}